\documentstyle[osa,manuscript]{revtex}

\newcommand{\be}{\begin{equation}}
\newcommand{\ee}{\end{equation}}
\begin{document}                
\title{Covariant gauge fixing and Kucha\v{r} decomposition}
\author{Petr H\'{a}j\'{\i}\v{c}ek}
\address{Institute for Theoretical Physics, University of Berne}
\author{Jerzy Kijowski}   
\address{Center for Theoretical Physics, Polish Academy of Sciences, Aleja
  Lotnik\'{o}v 32/46, 02-668 Warsaw, Poland}  
\maketitle
\begin{abstract}                
  The symplectic geometry of a broad class of generally covariant models is
  studied. The class is restricted so that the gauge group of the models
  coincides with the Bergmann-Komar group and the analysis can focus on the
  general covariance. A geometrical definition of gauge fixing at the
  constraint manifold is given; it is equivalent to a definition of a
  background (spacetime) manifold for each topological sector of a model.
  Every gauge fixing defines a decomposition of the constraint manifold into
  the physical phase space and the space of embeddings of the Cauchy manifold
  into the background manifold (Kucha\v{r} decomposition). Extensions of every
  gauge fixing and the associated Kucha\v{r} decomposition to a neighbourhood
  of the constraint manifold are shown to exist.
\end{abstract}


\section{Introduction}               
In 1971, Bergmann and Komar\cite{berg} wrote:
\begin{quote}
  ``...in general relativity the identity of a world point is not preserved
  under the theory's widest invariance group. This assertion forms the basis
  for the conjecture that some physical theory of the future may teach us how
  to dispense with world points as the ultimate constituents of space-time
  altogether.''
\end{quote}
We share this view and we are going to support it by revealing some of the
underlying mathematical structure.

The formulation of general relativity and, in fact, of any generally covariant
model, is based on the mathematical theory of (pseudo-)Riemannian manifolds.
There is, however, a catch: in the mathematics, even a naked manifold has
well-defined, distinguishable points. In the physics, points are defined and
distinguished {\em only} by values of physical fields or as positions of
physical objects. Attempts to take naked points seriously lead to well-known
paradoxes and problems. The first paradox of this kind was constructed by
Einstein (the `hole' argument \cite{hole}); a more recent example is due to
Fredenhagen and Haag \cite{F-H}. Any clean separation between spacetime points
on one hand and physical fields on the other violates the diffeomorphism
invariance (for an extended discussion of this point, see Stachel\cite{stach}
and Isham\cite{salamanca}). From the mathematical point of view, the space
that one works with is the space of geometries ${\mathrm Riem}{\mathcal
  M}/{\mathrm Diff}{\mathcal M}$ on a manifold $\mathcal M$ rather than the
space of metric fields ${\mathrm Riem} {\mathcal M}$ on the manifold $\mathcal
M$. In the space of geometries, points of the manifold ${\mathcal M}$ are
entangled with the metric fields and it is impossible to reconstruct
(disentangle) them in any natural, unique, way.

Accordingly, Einstein dynamics is not a field dynamics on any manifold. This
does not mean, however, that one cannot reduce it to such a field theory. For
example, the dynamics is reformulated as a system of partial differential
equations for some fields on a fixed background manifold in the study of the
Cauchy problem (see, e.g., a recent review\cite{CBYA}). This reduction is
based on choices of gauge (coordinate conditions). The choice of gauge plays,
in such a way, a two-fold role for generally covariant models: 1) it renders
the dynamics unique (as in any gauge field theory), and 2) it defines the
background manifold points. It is also well-known that the gauge group of such
models is much larger that just the diffeomorphism group of one fixed
manifold\cite{berg}.

The definition of background manifolds by means of gauge choices does not
violate the gauge invariance of the full theory, if one can show at the end
that the measurable results are independent of the choice; this has indeed
been possible for many problems of classical physics. Another popular method
of defining background manifolds is to expand certain sector of a given model
around a classical spacetime (such as, e.g., the Minkowski spacetime). A
special role given to a fixed classical spacetime enables one to use this
particular spacetime as a background, and to select the diffeomorphism group
of this spacetime as the remaining gauge symmetry. This is a strong
restriction of the original symmetry. The procedure might be justified, if
e.g.\ some kind of WKB approximation is valid in the situation considered and
the corresponding metric is a part of a classical solution from which the
iterative steps of the WKB method start.

In the present paper, we are going to study the symplectic geometry of quite a
general class of diffeomorphically invariant models. We shall concentrate on
those properties that are relevant to gauge fixing, gauge transformations, and
physical degrees of freedom. The main ideas are covariant gauge
fixing\cite{paris} and the Kucha\v{r} decomposition\cite{thirdway}; we shall
give a complete description of these ideas and their interconnection. The plan
of the paper is as follows.

In Sec.\ \ref{sec:exampl}, we analyze in some detail gauge choices using very
simple examples from general relativity. We try to separate the two aspects of
gauge fixing---the point definition and the coordinate choice---to motivate
our notion of covariant gauge fixing. We also briefly recapitulate Kucha\v{r}
``third way''\cite{thirdway}.

In Sec.\ \ref{sec:model}, we describe the properties of generally covariant
models that are needed for subsequent constructions. We present a list of
properties that can be considered as a kind of definition of the generally
covariant models. However, rather than attempting to identify a minimal set of
independent properties, we just collect all assumptions that will be
necessary for the proofs. For the sake of simplicity we also exclude all gauge
fields (such as Yang-Mills fields) so that we can focus on the issue of
general covariance.

Sec.\ \ref{sec:gauge} contains the constructions that are necessary for our
definition of covariant gauge fixing on the constraint manifold of the model.
The fixing identifies spacetime points belonging to different spacetime
solutions. In this way, a unique background manifold results and everything is
manifestly invariant with respect to coordinate transformations on this
manifold. The transformation between two covariant gauge fixings can be
described as a set of diffeomorphisms, one for each solution; such a set of
transformations is an element of the Bergmann-Komar group\cite{berg}. A
covariant gauge fixing is thus defined in a geometrical, coordinate free and
general manner. Still, it has a close relation to the usual way of choosing
gauge: a ``nice'' coordinate condition leads to a special case of such a
covariant gauge fixing.

The local existence of covariant gauge fixings is equivalent to the following
statement. For the sectors that are spatially compact, any open subset of the
generic part of the constraint surface on which the gauge fixing works is a
subset of a fiber bundle: its basis manifold is the physical phase space, its
typical fiber is the space of embeddings of the Cauchy surface into the
background manifold, and its group is that of diffeomorphisms of the
background. For the sectors that are not spatially compact, this description
is to be modified (see Sec.\ \ref{sec:gauge}). Each gauge fixing is equivalent
to a trivialization of this bundle, i.e.\ to a decomposition of the constraint
surface into a Cartesian product of the base and the typical fiber. Existence
of such decompositions has been first observed by Kucha\v{r}\cite{cylindr}; we
shall call them {\em Kucha\v{r} decompositions}.  In this way, we establish a
connection between covariant gauge fixings and Kucha\v{r} decompositions.

The main result of this paper is described in Sec.\ \ref{sec:ext}, where we
extend the Kucha\v{r} decomposition to a whole neighbourhood of the constraint
surface. The construction is based on the Darboux-Weinstein theorem\cite{wein}
and it shows explicitly that there are many such extensions.  As the
construction is based on an existence theorem, it will not be practically
viable in most cases of interest. However, Kucha\v{r} decompositions have as
yet been explicitly constructed only for very few cases, cylindrical
waves\cite{cylindr} and the Schwarzschild family\cite{schwarz}, and even the
question of existence was not clear. For most purposes (as, e.g., for
quantization), the explicit form of the decomposition outside the constraint
surface is not needed.

The mathematical language which is used in this paper and which enables
concise and effective formulations is that of vector bundles and symplectic
geometry of infinite dimensional manifolds modeled on Banach spaces (see,
e.g., Abraham, Marsden and Ratiu\cite{v-bundl}, Libermann and
Marle\cite{zielon} and Lang\cite{lang}). Typically, all these manifolds are
modeled on Sobolev spaces $H^s$ (see Marsden\cite{M}) but there is no
universal functional analytic framework for field theory: it seems that each
particular theory needs its own choice of the class of functions to which we
restrict our search of solution of field equations. Unfortunately, our results
cannot help to make this choice. Nevertheless, they are rigorous. What we
prove is the following statement: whenever a generally covariant field theory
is equipped with a {\em correct} functional analytic structure [``correct''
means that 1) the space of non-constraint Cauchy data is a Banach manifold, 2)
the constraint surface is its regular submanifold and 3) the gauge orbits form
a regular foliation of the latter] then this space is locally isomorphic to a
Cartesian product of the physical phase space and the cotangent bundle of
embeddings of the Cauchy surface into the background manifold. Each such local
isomorphism is connected with a covariant gauge fixing.

\section{Gauge in general relativity}
\label{sec:exampl}
In this section we analyze the gauge choice in general relativity and
review the original Kucha\v{r} decomposition.

To discuss the gauge choice, we use a strongly simplified model. This will
motivate our subsequent definitions and constructions.

Consider the Schwarzschild solutions to the Einstein equations in the future
of the influence (white hole) horizon. They form a one-dimensional family and
the value of the Schwarzschild mass $M\in (0,\infty)$ distinguishes different
elements of the family from each other. The metric can be given the form
\be
  ds^2 = -\left(1-\frac{2M}{R}\right)dW^2 + 2dW\,dR + R^2\,ds_2^2,
\label{a.1}
\ee
where $ds_2^2$ is the metric of a 2-sphere of radius 1; $W$ and $R$ are the
advanced Eddington-Finkelstein coordinates with the domains
\be
  R \in (0,\infty),\quad W\in (-\infty,\infty).
\label{b.1}
\ee
Nothing seems to prevent us from considering Eq. (\ref{a.1}) as a
one-dimensional {\em set of metric fields} on a fixed background manifold
${\mathcal M}_1 = {\mathbf R}^2\times S^2$ in the coordinate chart $W$, $R$,
$\vartheta$ and $\varphi$ [of course, at least two charts
$(\vartheta,\varphi)$ and $(\vartheta',\varphi')$ are necessary to cover
$S^2$]. The same metric can, however, also be given another form, if we
pass to the Kruskal coordinates $U$, $V$, $\vartheta$ and $\varphi$. 
\be
  ds^2 = -\frac{16M^2}{\kappa(-UV)}
  e^{-\kappa(-UV)}\,dU\,dV + 4M^2\kappa^2(-UV)\,ds_2^2,
\label{b.2}
\ee
where $\kappa : (-1,\infty) \mapsto (0,\infty)$ is the well-known Kruskal
function defined by its inverse, $\kappa^{-1}(x) = (x - 1)e^x$ for 
$x\in(0,\infty)$; the coordinates $U$ and $V$ are restricted to the domains
\be
  U\in (-\infty,\infty),\quad V\in (0,\infty),
\label{c.1}
\ee 
in order that the same parts of the spacetimes as given by Eq.\ (\ref{b.1})
are covered. 

Let us look carefully at the transformation between the Eddington-Finkelstein
and Kruskal coordinates:
\be
  U = \left(\frac{R}{2M} -
    1\right)\exp\left(\frac{R}{2M}\right)\exp\left(-\frac{W}{4M}\right),\quad
  V = \exp\left(\frac{W}{4M}\right) 
\label{c.23}
\ee
(the transformation of the angular coordinates is trivial). Eqs.\ (\ref{c.23})
do not represent any coordinate transformation on ${\mathcal M}_1$, because
they are solution dependent: the right-hand sides are non-trivial functions of
$M$. They can only be interpreted as coordinate transformations, if we view
Eq.\ (\ref{a.1}) together with the manifold ${\mathcal M}_1$ as a {\em family
  of solutions} $\{({\mathcal M}_1,g_M)\}$ rather that a family of metric
fields $\{g_M\}$ on a background manifold ${\mathcal M}_1$.

Eqs.\ (\ref{a.1}) and (\ref{b.1}) express the Schwarzschild family in two
different gauges. We can see from the above that a gauge transformation in
general relativity is a {\em set of coordinate transformations}, one
transformation for each solution (cf.\ Bergmann and Komar\cite{berg}), rather
than a coordinate transformation on one manifold.

The illusion of a background manifold only arises, if one
pastes together all solution manifolds in such a way that points with the same
value of coordinates representing some gauge are considered to be
identical. Thus, the background manifold ${\mathcal M}_1$ results, if we
identify all points that have the same values of the Eddington-Finkelstein
coordinates $W$ and $R$; an analogous background manifold ${\mathcal M}_2$ for
metrics (\ref{b.2}) is constructed by identifying all points with the same
values of Kruskal coordinates $U$ and $V$. It should be clear that, in spite
of the fact that both manifolds are formally diffeomorphic to each other, they
nevertheless represent two very different localizations of geometrical
properties of the Schwarzschild family. For example, the position of the event
horizon on ${\mathcal M}_1$, which is given by the equation $R = 2M$, is not
well-defined (fuzzy): the horizon of each solution lies at different points of
${\mathcal M}_1$. On the contrary, the position of the event horizon on
${\mathcal M}_2$, which is given by $U = 0$, is well-defined (sharp), because
it is solution independent.

This all is well-known and rather trivial. Still, the detailed form of the
above analysis simplifies the understanding of the following point: a choice
of gauge in general relativity mixes two different things: 1) It defines how
points of different solution manifolds are to be identified so that a
background manifold can be constructed. 2) It chooses definite coordinates on
the background manifold. It is already the first step alone that delivers what
we require from a gauge fixing: a {\em unique} metric field on a background
manifold, say ${\mathcal M}$ for any solution (determined by the value of $M$
in our example). This metric field can be given in any coordinate system on
the background manifold; that is, all coordinate transformations on ${\mathcal
  M}$ are allowed (these should be $M$-independent for our example).
Everything can be made manifestly covariant with respect to such
transformations, in spite of the clear fact that the gauge has been fixed.

We do not know if this observation has ever been put forward in its full
generality, but it surely has been done for the perturbative approach to
general relativity by DeWitt\cite{DW}. Let us explain DeWitt idea in more
detail in order to prevent misunderstanding. DeWitt chooses a particular
classical {\em background spacetime} $({\mathcal M},g)$ (his method has, in
fact, been called {\em background field method}). All other spacetimes in some
vicinity of $({\mathcal M},g)$ are described by small disturbances $\delta g$
around $g$. Two kinds of gauge fixing is now possible: The first kind is just
a choice of coordinates $x^\mu$ on $\mathcal M$; with respect to $x^\mu$, the
metric $g$ and the disturbance $\delta g$ have components $g_{\mu\nu}(x)$ and
$\delta g_{\mu\nu}(x)$ and they transform as two tensor fields with respect to
changes of these coordinates. A gauge transformation of the second kind is a
small diffeomorphism $\delta\xi$ on $\mathcal M$. The background metric field
changes then by a Lie derivative ${\mathcal L}_\xi g$; in coordinates,
$g_{\mu\nu}(x) \mapsto g_{\mu\nu}(x) + \delta\xi_{\mu;\nu} +
\delta\xi_{\nu;\mu}$, where the semicolon denotes the covariant derivative
defined by the metric $g$. Such a change is not considered as a change of a
physical state, so disturbances of the form $\delta g_{\mu\nu} =
\delta\xi_{\mu;\nu} + \delta\xi_{\nu;\mu}$ for any small vector field
$\delta\xi$ on $\mathcal M$ are considered as ``pure gauges''. To fix a gauge
of second kind, DeWitt requires a {\em supplementary condition} that is
covariant with respect to coordinate transformations on $\mathcal M$ (gauge of
the first kind). Thus, the background field method becomes covariant; even the
field equations after the gauge of the second kind has been fixed are
covariant in his formalism.

DeWitt's supplementary condition hinders gauge transformations of the second
kind; these form a group, namely the group Diff$\mathcal M$ of diffeomorphisms
of the background manifold $\mathcal M$. On the other hand, the gauge
transformations that we are considering form the much larger Bergmann-Komar
group\cite{berg}. Thus, there is an analogy, but not complete equivalence
between the two ideas of gauge fixing.

Covariant gauge fixing is connected\cite{paris} to an idea due to
Kucha\v{r}\cite{thirdway}. Let us describe this briefly in the rest of this
section (for more details, see Kucha\v{r}\cite{canada,goa}).

The Hamiltonian formalism for general relativity has been described in an
elegant 3-covariant form by Arnowitt, Deser and Misner\cite{ADM}. The action 
depends on the ADM variables $g_{kl}(x)$ and $\pi^{kl}(x)$ as follows
\[
  S = \int dt\int_\Sigma d^3x\Bigl(\pi^{kl}(x)\dot{g}_{kl}(x) - {\mathcal
  N}(x){\mathcal H}(x) - {\mathcal N}^k(x){\mathcal H}_k(x)\Bigr),
\]
where $\Sigma$ is a three-dimensional manifold, ${\mathcal
  H}[g_{kl},\pi_{kl};x)$ and ${\mathcal H}_k[g_{kl},\pi_{kl};x)$ are the
constraints [functionals of $g_{kl}(x)$ and $\pi^{kl}(x)$ and functions of
$x$], and ${\mathcal N}(x)$ and ${\mathcal N}^k(x)$ are Lagrange
multipliers\cite{ADM}.

Kucha\v{r} observed that one can sometimes make a canonical transformation,
\be
 g_{kl}(x), \pi^{kl}(x)  \mapsto X^\mu(x), P_\mu(x), q^\alpha(x), p_\alpha(x)
\label{h.1}
\ee
so that the action acquires the form
\[
  S = \int dt\int_\Sigma d^3x\left(p_\alpha(x)\dot{q}^\alpha(x) +
  P_\mu(x)\dot{X}^\mu(x) - {\mathcal N}^\mu(x){\mathcal H}_\mu(x)\right),
\]
where ${\mathcal H}_\mu(x)$ are linear combinations of the original
constraints ${\mathcal H}(x)$ and ${\mathcal H}_k(x)$. The new constraints
read ${\mathcal H}_\mu(x) = P_\mu(x) + H_\mu[X,q,p;x)$, where $H_\mu[X,q,p;x)$
are ``true Hamiltonians''.

The variables $X^\mu(x)$ describe embeddings of the three-dimensional Cauchy
surface $\Sigma$ with coordinates $x^k$ into a four-dimensional background
manifold $\mathcal M$ with coordinates $X^\mu$. The function $\int_\Sigma
d^3x{\mathcal N}^\mu(x){\mathcal H}_\mu(x)d\epsilon$ generates an
infinitesimal canonical transformation in the phase space that describes the
dynamical evolution from the slice defined by the embedding $X^\mu(x)$ to the
slice defined by the embedding $X^\mu(x) + {\mathcal N}^\mu(x)d\epsilon$. In
this way, the dynamics is made completely independent from any additional
structure on $\mathcal M$ such as a particular foliation.

If a transformation (\ref{h.1}) exists, one can go a step further and pass to
what Kucha\v{r}\cite{goa} called the Heisenberg picture (see also
Kijowski\cite{kijow}). This is another canonical transformation,
\be
  X^\mu(x),P_\mu(x),q^\alpha(x),p_\alpha(x) \mapsto X^\mu(x),{\mathcal
  H}_\mu(x),q_0^\alpha(x),p_{0\alpha}(x),
\label{KD}
\ee
where $q_0^\alpha(x)$ and $p_{0\alpha}(x)$ are values of $q^\alpha(x)$ and
$p_\alpha(x)$ at some particular embedding $X^\mu_0(x)$. Clearly,
$q_0^\alpha(x)$ and $p_{0\alpha}(x)$ are constants of motion:
\[
  \{q_0^\alpha(x),{\mathcal H}_\mu(y)\} = \{p_{0\alpha}(x),{\mathcal
  H}_\mu(x)\} = 0,
\]
so that the action becomes
\be
  S = \int dt\int_\Sigma d^3x\left(p_{0\alpha}(x)\dot{q}_0^\alpha(x) +
    {\mathcal H}_\mu(x)\dot{X}^\mu(x) - {\mathcal N}^\mu(x) 
    {\mathcal H}_\mu(x) \right);
\label{kuch-a}
\ee
this {\em is} a special case of the form of the action after the first
transformation, but the true Hamiltonians are zero and the $P$'s are identical
to the constraints now. It is the transformations (\ref{h.1}) and (\ref{KD})
and the corresponding variables that we shall call {\em Kucha\v{r}
  decomposition}.

It is clear that Kucha\v{r} decomposition must implicitly include a gauge
fixing not only because it leads to a well-defined background manifold
$\mathcal M$, but also to a fixed coordinate system $X^\mu$ on it. Indeed,
Kucha\v{r} decomposition also defines a particular set of metric fields on
$\mathcal M$ by one of the canonical transformation equations, namely that of
the form
\[
  g_{kl}(x) = g_{\mu\nu}\Bigl(q(x),p(x),X(x)\Bigr)X^\mu_{,k}(x)X^\nu_{,l}(x)
\]
for any embedding $X^\mu(x)$, where $g_{\mu\nu}(q,p,X)$ is a metric field for
any value of the variables $q^\alpha(x)$ and $p_\alpha(x)$ (see, e.g.\ 
Kucha\v{r}, Romano and Varadarajan\cite{KRV}). The KRV metric
$g_{\mu\nu}(q,p,X)$ is clearly an analog of the metric (\ref{a.1}) [or
(\ref{b.2})]: $q$ and $p$ play the role of the Schwarzschild mass $M$, and $X$
that of the Eddington-Finkelstein (Kruskal) coordinates.

In the present paper, we shall describe Kucha\v{r} decomposition in geometric
(that is, coordinate-free) terms.

\section{The generally covariant models}
\label{sec:model}
We shall consider a class of constrained dynamical systems that are in certain
respects similar to general relativity. As examples, general relativity,
possibly coupled to matter fields, 2+1 gravity\cite{carlip}, possibly with
particle-like sources, and the spherically symmetric gravitating thin
shell\cite{PH1} can be mentioned. In this section, we define the class by a
list of properties. For some of the models named above, not all of these
properties have been fully established yet.

\subsection{The form of dynamical trajectories}
\label{sec:A}
A dynamical trajectory---or classical solution---of each such model consists
of two parts. The first part is a spacetime $({\mathcal M},g)$, where 
${\mathcal M}$ is a manifold of dimension $D$ and $g$ is a metric of
(Lorentzian) signature $D-2$. Each such spacetime will be called a {\em
  solution spacetime}. Second, any dynamical trajectory may contain
additional fields and branes (submanifolds of $\mathcal M$ carrying
other fields---they play the role of trajectories of particles, strings,
shells etc.) on $\mathcal M$, which we shall describe by the symbol $\phi$;
thus a dynamical trajectory can be denoted by $({\mathcal M},g,\phi)$. Just
for the sake of simplicity, we assume that there are no gauge fields within
$\phi$, but this restriction can be removed easily.

\subsection{Diffeomorphism invariance}
\label{sec:B}
The dynamical equations of each such model are generally covariant\cite{foot}
with respect to all coordinate transformations on $\mathcal M$. This implies
that any system $g$ and $\phi$ of fields and branes satisfying the
dynamical equations on a manifold $\mathcal M$ can be pushed forward by any
diffeomorphism $\varphi \in {\mathrm Diff}{\mathcal M}$ to a different set
$\varphi_*g$ and $\varphi_*\phi$ on $\mathcal M$, which also satisfies the
dynamical equations. Indeed, if $X$ are any coordinates on $\mathcal M$, and 
$g(X)$ and $\phi(X)$ the components of all fields and branes with respect to
$X$, then $\varphi_*g$ and $\varphi_*\phi$ have exactly the same components
with respect to the pushed-forward coordinates $X' := X\circ\varphi^{-1}$.
They satisfy, therefore, the dynamical equations of exactly the same form.
Observe that even the spinor fields can be pushed forward in this way, because
the metric is, so the push-forward of any $D$-frame that is orthonormal with
respect to the metric $g$ will be orthonormal with respect to $\varphi_*g$.

Hence, if $({\cal M},g,\phi )$ is a dynamical trajectory, then $({\cal
  M},\varphi_* g,\varphi_* \phi )$ is also one for any $\varphi \in $
Diff${\cal M}$. This feature is called {\em diffeomorphism invariance}.  In
general, the set $(\varphi_* g,\varphi_* \phi )$ of fields and branes on
${\mathcal M}$ is different from the set $(g,\phi )$. However, we are going to
treat them as physically equivalent if only $\varphi \in \mbox{\rm
  Diff}_\infty{\cal M}$, where Diff$_\infty{\mathcal M}$ is a subgroup of
Diff${\mathcal M}$ composed of those diffeomorphisms that are ``trivial at
infinity''. For example, if the solution spacetime is asymptotically flat, the
elements of Diff$_\infty{\mathcal M}$ must move neither the points at the
infinity nor the frames at these points. For $\mathcal M$ spatially compact,
there is no ``infinity'' and Diff$_\infty{\mathcal M}$ simply coincides with
the entire diffeomorphism group Diff${\mathcal M}$. Thus, the physical state
of the system under consideration is always described by a whole class of
equivalent trajectories {\em modulo} the action of the group $\mbox{\rm
  Diff}_\infty{\cal M}$. We denote such a class $\{ ({\cal M},g,\phi )\} $,
where $({\cal M},g,\phi )$ is a particular set of fields and branes on ${\cal
  M}$ satisfying the dynamical equations.

Even if the whole group $\mbox{\rm Diff}{\cal M}$ (i.e., also those
diffeomorphisms which are non-trivial ``at infinity'') forms the symmetry
group of the theory, the gauge group of the model will be constructed only
from the subgroup $\mbox{\rm Diff}_\infty{\cal M}$. The reason for this
decision is obvious if we think e.g.\ about special-relativistic mechanics of
a free particle: considering two motions as being physically equivalent if
they only differ by the action of an element of a symmetry group (in that case
it would be the Poincar\'e group) would be an abuse of the very notion of
symmetry in physics. The physical phase space resulting from such a
construction would consist of a single equivalence class, composed of all
possible physical situations. In such a zero-dimensional space no non-trivial
dynamics is possible.

\subsection{The initial data}
\label{sec:C}
We assume further that each model determines a class of $(D-1)$-dimensional
manifolds; each such manifold $\Sigma$ is called {\em initial manifold}.
Further, it also determines a class of a system of some fields and membranes
$\gamma$, $\psi$, $\dot{\gamma}$ and $\dot{\phi}$ on $\Sigma$. The object
$(\Sigma,\gamma,\psi,\dot{\gamma},\dot{\psi})$ built up from the elements of
the classes is then called an {\em initial datum} of the model. For example,
in general relativity, any three-dimensional Riemannian manifold can serve as
an initial manifold; the field $\gamma$ is a Riemann metric on it,
$\dot{\gamma}$ is a symmetric tensor field $K_{kl}$ on $\Sigma$ and there is
no $\psi$ and $\dot{\psi}$.

An important connection of initial data to the dynamical trajectories is the
following. Let $({\mathcal M},g,\phi)$ be a dynamical trajectory; then,
$(D-1)$-dimensional submanifolds of ${\mathcal M}$ satisfying certain
requirements are called {\em Cauchy surfaces} in ${\mathcal M}$. Each Cauchy
surface is a possible initial manifold from the class above. Let $\Sigma$ be a
Cauchy surface for the dynamical trajectory $({\mathcal M},g,\phi)$; then the
fields and branes $g$ and $\phi$ determine a unique initial datum
$(\Sigma,\gamma,\psi,\dot{\gamma},\dot{\psi})$. For example, $\gamma$ and
$\psi$ are the pull-backs of the fields to, and intersections of the branes
with, the surface $\Sigma$; $\dot{\gamma}$ and $\dot{\phi}$ are some
geometrical quantities on $\Sigma$ constructed from the fields and their first
derivatives at $\Sigma$, and from projections into $\Sigma$ of the normalized
$D$-velocities of the branes. Such an initial datum is called {\em induced} on
the Cauchy surface $\Sigma$ by the dynamical trajectory $({\mathcal
  M},g,\phi)$.  Then the dynamical equations (and the asymptotic conditions
for $g$ and $\phi$) of the model imply some asymptotic conditions for
$\gamma$, $\psi$, $\dot{\gamma}$ and $\dot{\phi}$ and some relations between
them on $\Sigma$ that are called {\em constraints}.

\subsection{The existence and uniqueness of dynamical trajectories}
\label{sec:D}
We assume further that the dynamical equations and all initial data have the
following property. For each initial datum
$(\Sigma,\gamma,\psi,\dot{\gamma},\dot{\psi})$ that satisfies the constraints
there is a {\em unique} Diff$_\infty(\Sigma\times{\mathbf R})$-class
$\{(\Sigma\times{\mathbf R},g,\phi)\}$ of maximal dynamical trajectories such
that each element $(\Sigma\times{\mathbf R},g,\phi)$ of the class contains a
Cauchy surface on which the induced datum is diffeomorphic to
$(\Sigma,\gamma,\psi,\dot{\gamma},\dot{\psi})$. This implies that the set of
objects defining initial data must be complete in a certain sense.

The uniqueness of the maximal dynamical trajectory is understood in the sense
of Choquet-Bruhat and Geroch\cite{C-G}. It has been shown for general
relativity that the solution spacetimes of maximal dynamical trajectories are
globally hyperbolic; we will assume the same property for all our models.
According to a theorem of Geroch\cite{ger}, each globally hyperbolic spacetime
in general relativity can be completely foliated by spacelike hypersurfaces,
each of them being diffeomorphic to $\Sigma$. This leads us to assume that
${\mathcal M} = \Sigma\times{\mathbf R}$.

The uniqueness implies the following property of the dynamical equations.
Suppose that $({\mathcal M},g,\phi)$ is a maximal dynamical trajectory for the
initial datum $(\Sigma,\gamma,\psi,\dot{\gamma},\dot{\psi})$ (so
$(\Sigma,\gamma,\psi,\dot{\gamma},\dot{\psi})$ satisfies the constraints and
the asymptotic conditions). Let $\Sigma'$ be an arbitrary Cauchy surface in
${\mathcal M}$, and the initial datum
$(\Sigma',\gamma',\psi',\dot{\gamma}',\dot{\psi}')$ be induced by the
dynamical trajectory $({\mathcal M},g,\phi)$ on $\Sigma'$. Then
$(\Sigma',\gamma',\psi',\dot{\gamma}',\dot{\psi}')$ satisfies the constraints
and the asymptotic conditions, and any representative $({\mathcal
  M},g',\phi')$ of the unique maximal dynamical trajectory corresponding to
$(\Sigma',\gamma',\psi',\dot{\gamma}',\dot{\psi}')$ is diffeomorphic to
$({\mathcal M},g,\phi)$.

\subsection{The phase space}
\label{sec:E}
Let us consider the set $\Gamma'_1$ of all initial data, and the subset
$\Gamma_1$ of those data that satisfy the constraints (and asymptotic
conditions). In the case of spatially compact sectors, we exclude all data
from $\Gamma'_1$ that lie in $\Gamma_1$ and determine maximal dynamical
trajectories admitting any symmetry (or any symmetry that is higher than the
symmetry following from the definition of the model). It seems that this
deleting is not necessary\cite{M} for sectors that are not spatially compact.
Thus, not only (global) Killing vector fields in the solution spacetimes
$({\mathcal M},g)$ are forbidden, but also any finite (discrete) symmetry. Let
us denote the resulting sets by $\Gamma'_2$ and $\Gamma_2$. We assume that a
subset $\Gamma'$ of $\Gamma'_2$ has been organized in such a way that
$\Gamma'$ is a manifold modeled on a Banach space and $\Gamma := \Gamma' \cap
\Gamma_2$ is a closed submanifold of $\Gamma'$. In general, $\Gamma'$ is an
open submanifold of the phase space of the model. In general relativity,
$\Gamma$ was shown to be a $C^\infty$-submanifold (for reviews, see Fischer \&
Marden\cite{F-M} and Marsden\cite{M}) The condition of no symmetry (even a
discrete one) for the spatially compact sectors is necessary for the
construction of the covariant gauge fixing in the next section to work.

We assume that each model defines a symplectic form $\Omega'$ on $\Gamma'$
(this may be determined by the variational principle of the model under
study---for more discussion, see Kijowski and Tulczyjew\cite{K-T}). For
example, in general relativity, $\Omega' = d\Theta'$ and $\Theta' =
\int_\Sigma d^3x\, \pi^{kl}(x)d\gamma_{kl}(x)$, where $\pi^{kl} :=
\text{Det}(\gamma_{mn})^{1/2} (\gamma^{kl}\gamma^{ij} -
\gamma^{ki}\gamma^{lj})K_{ij}$ is a super-local functional of $\gamma_{kl}$
and $K_{kl}$. We assume further that $\Gamma$ is a coisotropic\cite{zielon}
submanifold of $\Gamma'$ with respect to $\Omega'$.  That is the following
property. Let $p\in \Gamma$ and let $L_p(\Gamma)$ be a subspace of
$T_p(\Gamma')$ defined by
\be 
  L_p(\Gamma) := \{v\in T_p(\Gamma')|\Omega'(v,u)
  = 0 \quad \forall u\in T_p(\Gamma)\}.
\label{char}
\ee
One can write alternatively\cite{zielon} $L_p(\Gamma) = {\mathrm
  orth}_{\Omega'}T_p(\Gamma)$. $\Gamma$ is coisotropic if $L_p(\Gamma)
\subset T_p(\Gamma)$. Hence, the pull-back $\Omega$ of $\Omega'$ to $\Gamma$
is a presymplectic form on $\Gamma$. The space $L_p(\Gamma)$ is called the
{\em characteristic space} of $\Omega$.
  
The subbundle $L(\Gamma) := \{\Bigl(p,L_p(\Gamma)\Bigr)|p\in \Gamma\}$ of the
tangent bundle $T(\Gamma)$ is an integrable subbundle, because $\Omega$ is
closed; it is called the {\em characteristic bundle} of $\Omega$. Let us call
the maximal integral manifolds of $L(\Gamma)$ {\em c-orbits}.

In the case of infinite-dimensional models, additional, model-specific
assumptions\cite{M,F-M} are needed for the proofs that $\Gamma$ is a
submanifold and that c-orbits with suitable properties exist.

\subsection{Relation between the c-orbits and the maximal dynamical
trajectories} 
\label{sec:F} 
We assume that there is a relation between c-orbits and dynamical trajectories
of the model as follows.  Let $o$ be a c-orbit and $p\in o$. Let $p$ be the
initial datum $(\Sigma,\gamma,\psi,\dot{\gamma},\dot{\psi})$ and let
$({\mathcal M},g,\phi)$ be a maximal development of
$(\Sigma,\gamma,\psi,\dot{\gamma},\dot{\psi})$.  Then the initial datum
$(\Sigma',\gamma',\psi',\dot{\gamma}',\dot{\psi}')$ corresponding to any point
$q\in o$ defines a unique Cauchy surface $\Sigma'$ in ${\mathcal M}$, by the
condition that the dynamical trajectory $({\mathcal M},g,\phi)$ induces the
initial datum $(\Sigma',\gamma',\psi',\dot{\gamma}',\dot{\psi}')$ on
$\Sigma'$. Moreover, $\Sigma'$ can be obtained from $\Sigma$ by the action of
some $\varphi\in\text{Diff}_\infty{\mathcal M}$. In general, all Cauchy
surfaces that correspond to points in the c-orbit $o$ form an open subset of
$\mathcal M$.

If the Cauchy surfaces are not compact, then they must satisfy certain
boundary conditions. For example, in the case of asymptotically flat solution
spacetimes, the Cauchy surfaces must be asymptotically flat and all {\em
  coincide} with each other at infinity in order to define points of the same
c-orbit in $\Gamma$.  We can also say: let $p$ and $q$ be any two points from
the same c-orbit $o$.  Let the corresponding initial data determine dynamical
trajectories $({\mathcal M},g,\phi)$ and $({\mathcal M'},g',\phi')$.  Then
these two dynamical trajectories are diffeomorphic to each other. In this way,
any c-orbit $o$ determines a class of Diff$_\infty{\mathcal M}$-equivalent
maximal dynamical trajectories.

Observe that a dynamical trajectory $({\mathcal M},g,\phi)$ corresponding to
the datum $(\Sigma,\gamma,\psi,\dot{\gamma},\dot{\psi})$ contains {\em
  exactly} one surface $\Sigma\subset {\mathcal M}$ such that the datum
induced on $\Sigma$ coincides with
$(\Sigma,\gamma,\psi,\dot{\gamma},\dot{\psi})$. Indeed, two different Cauchy
surfaces carrying initial data that are Diff$_\infty{\mathcal M}$-equivalent
to each other would imply existence of a non-trivial symmetry of the dynamical
trajectory $({\mathcal M},g,\phi)$, and such points have been excluded from
$\Gamma$.

\subsection{The physical phase space}
\label{sec:G}
The last important property we assume is that the set of the c-orbits in the
constraint surface form a {\em quotient manifold}\cite{lang} $\Gamma/o$ with
the natural projection $\pi : \Gamma \mapsto \Gamma/o$ being a {\em
  submersion}\cite{lang}. Furthermore, there is a unique symplectic form
$\tilde{\Omega}$ on $\Gamma/o$ such that $\Omega = \pi^*\tilde{\Omega}$, where
$\pi^*$ is the pull-back of forms by $\pi$. Our reduced symplectic space
$(\Gamma/o,\tilde{\Omega})$ is, in general, an open subset of the {\em
  physical phase space}. For example, in general relativity, some aspects of
the physical phase space are discussed by Marsden\cite{M} and Fischer and
Moncrief\cite{F-Mo}.

We maintain that all information about physical properties of the models is
contained in the physical phase spaces. One is, however, forced to use the
extended structures $\Gamma$ and $\Gamma'$, because it is often difficult in
practice to perform the reduction to the physical phase space and to find an
explicit parametrization of it.

\section{Covariant gauge fixings}
\label{sec:gauge}
The general structure described in the previous section enables us to work out
a geometric definition of gauge fixing based on the ideas of the previous
paper\cite{paris}. This will concern only the diffeomorphism group---as
explained, we assume that there are no other gauge groups acting.

Let $o$ be an arbitrary c-orbit, $\Sigma_o$ be the manifold structure of the
corresponding Cauchy surfaces and $\{({\mathcal M},g,\phi)\}_o$ the
diffeomorphism class of the maximal dynamical trajectories determined by $o$.
Let us choose one fixed representative $({\mathcal M}_o,g_o,\phi_o)$ from this
class for each $o$. Consider the set Emb$_c(\Sigma_o,{\mathcal M}_o)$ of all
embeddings of $\Sigma_o$ in ${\mathcal M}_o$ such that the embedded
submanifold is a Cauchy surface; we call such embeddings {\em Cauchy
  embeddings}. If $\Sigma_o$ is not compact, we restrict the space
Emb$_c(\Sigma_o,{\mathcal M}_o)$ to a class of embeddings that satisfy the
boundary conditions formulated in subsection \ref{sec:F}.  We assume that
Emb$_c(\Sigma_o,{\mathcal M}_o)$ is an open subset of the space
Emb$(\Sigma_o,{\mathcal M}_o)$ of all (smooth) embeddings of $\Sigma_o$ in
${\mathcal M}_o$ (that also satisfy the boundary conditions in the non-compact
case). Discussion of this point for the compact cases is given in Isham and
Kucha\v{r}\cite{I-K}. Then it follows from the assumptions in Sec.\ 
\ref{sec:F} that there is an injection
\[
  \rho_o: o \mapsto {\mathrm Emb}_c(\Sigma_o,{\mathcal M}_o)
\]
such that each point $p$ of $o$ is mapped onto that Cauchy surface
$h(\Sigma_o)$ in ${\mathcal M}_o$, $h\in{\mathrm Emb}_c(\Sigma_o,{\mathcal
  M}_o)$, on which the initial datum $p$ is induced by $({\mathcal
  M}_o,g_o,\phi_o)$. Such a map $\rho_o$ is not unique. It depends on the
chosen representative $({\mathcal M}_o,g_o,\phi_o)$, but any two possible
$\rho_o$'s differ by a diffeomorphism $\varphi \in \text{Diff}_\infty{\mathcal
  M}_o$, $\rho'_o = \varphi\circ\rho_o$.

We assume that $\Gamma'$ and Emb$_c(\Sigma_o,{\mathcal M}_o)$ has been given
differentiable structure such that the map $\rho_o$ together with its inverse
become differentiable.  This implies the following properties. Let $h :
\Sigma_o \mapsto {\mathcal M}_o$ be a point of Emb$_c(\Sigma_o,{\mathcal
  M}_o)$ that lies in the range of $\rho_o$. Then the elements of the tangent
space $T_h$Emb$_c(\Sigma_o,{\mathcal M}_o)$ to Emb$_c(\Sigma_o,{\mathcal
  M}_o)$ at $h$ are vector fields $V(x)$ along $h(\Sigma_o)$ in ${\mathcal
  M}_o$, where $x \in h(\Sigma_o)$ (they may have to satisfy some suitable
smoothness and boundary conditions\cite{I-K}); all such vector fields form the
tangent space $T_h{\mathrm Emb}_c(\Sigma_o,{\mathcal M}_o)$.  In particular,
the following conditions are to be satisfied:
\begin{enumerate}
\item There is a smooth family of smooth curves $C_x(\lambda)$ in
  ${\mathcal M}_o$ such that $C_x(0) = x$ and $C'_x(0) = V(x)$ for each $x \in
  h(\Sigma_o)$. 
\item There is an $\epsilon > 0$ such that $C_x(\lambda)$ is well-defined for
    each $x \in h(\Sigma_o)$ and each $\lambda$ such that $|\lambda| <
    \epsilon$. 
\item For any fixed $\lambda$ such that $|\lambda| < \epsilon$, the
    expression $C_x(\lambda)$ defines a map $c_\lambda : x \mapsto {\mathcal
      M}_o$ for all $x\in h(\Sigma_o)$ by $c_\lambda(x) = C_x(\lambda)$;
    then we require that $c_\lambda\circ h : \Sigma_o \mapsto {\mathcal
      M}_o$ belongs to Emb$_c(\Sigma_o,{\mathcal M}_o)$ for all $\lambda$, or
    $c_\lambda\Bigl(h(\Sigma_o)\Bigr)$ is a Cauchy surface in ${\mathcal M}_o$
    for all $\lambda \in (-\epsilon,\epsilon)$.
\end{enumerate}
If there is such a family, then there are many.

Consider the map $\rho^{-1}_o\biggl(c_\lambda\Bigl(h(\Sigma_o)\Bigr)\biggr) :
\lambda \mapsto o$.  It defines a curve through the point $p =
\rho^{-1}_o\Bigl(h(\Sigma_o)\Bigr)$ in $o$. The differentiability requirements
on $\rho_o$ mean that it is a smooth curve in $o$ with a well-defined
tangential vector $v$ at $p \in o$; $v$ is non-zero if $V(x) \neq 0$, is
tangential to $o$ and depends only on the vector field $V(x)$, not on a
particular family of curves $C_x(\lambda)$.

In the opposite direction, let $p\in o$ and $v$ be a vector at $p$ tangential
to $o$. Then there is a curve $\tilde{C}(\lambda)$ in $o$ for $|\lambda| <
\epsilon$, $\epsilon > 0$, such that $\tilde{C}(0) = p$ and $\tilde{C}'(0) =
v$ and the map $\rho_o$ determines a family of Cauchy surfaces
$\{\Sigma_\lambda\}$ in ${\mathcal M}_o$ by $\Sigma_\lambda =
\rho_o\Bigl(\tilde{C}(\lambda)\Bigr)(\Sigma_o)$.  In particular, any fixed
point $x\in \Sigma_o$ will be mapped by the Cauchy embedding
$\rho_o\Bigl(\tilde{C}(\lambda)\Bigr)$ to a point that we denote by
$C_x(\lambda)$ in a neighbourhood of $\rho_o(p)(x)$. These curves have tangent
vectors for each $x$ at $\lambda=0$ that we denote by $V(x)$; hence,
\[
  V(x) = \biggl(\rho_o\Bigl(C(\lambda)\Bigr)(x)\biggr)'|_{\lambda=0}.
\]
Again, $V(x)$ must be from $T_{\rho_o(p)}{\mathrm Emb}(\Sigma_o,{\mathcal
  M}_o)$, non-zero if $v \neq 0$ and independent of a particular curve
$\tilde{C}(\lambda)$ chosen in $o$. For example, in general relativity, this
differentiability has been shown by Moncrief\cite{mon-dif}.

Consider the submanifold $\Gamma_\Sigma$ containing all points of $\Gamma$
that correspond to a fixed initial manifold $\Sigma$. As a rule, different
$\Gamma_\Sigma$'s are topologically separated in $\Gamma$, and can be called
(topological) {\em sectors} of the model. A {\em covariant gauge fixing} in
$\Gamma_\Sigma$ is the set of maps $\{\iota_o\}$ that satisfies two
requirements:
\begin{quote}
  1. For each c-orbit $o \in \Gamma_\Sigma/o$, $\iota_o : {\mathcal M}_o
  \mapsto \Sigma\times{\mathbf R}$ is a well-defined differentiable 
  injection with differentiable inverse. 
\end{quote}
Any such map induces a differentiable injection $\tilde{\iota}_o :
\text{Emb}(\Sigma,{\mathcal M}_o) \mapsto
\text{Emb}(\Sigma,\Sigma\times{\mathbf R})$ with differentiable inverse by
$\tilde{\iota}_oh := \iota_o\circ h$ for any $h\in \text{Emb}(\Sigma,{\mathcal
  M}_o)$. Define $\sigma : \Gamma_\Sigma \mapsto {\mathrm
  Emb}(\Sigma,\Sigma\times{\mathbf R})$ by $\sigma|_o :=
\tilde{\iota}_o\circ\rho_o$ for all $o\in\Gamma_\Sigma/o$. Then
\begin{quote}
  2.  $\sigma$ is a differentiable map of $\Gamma_\Sigma$ into
  $\text{Emb}(\Sigma,\Sigma\times{\mathbf R})$.
\end{quote}

Observe that the maps $\rho_o$ are arbitrary and are not supposed to have any
relations for different $o$'s so that the dynamical trajectories $({\mathcal
  M}_o,g_o,\phi_o)$ and $({\mathcal M}_{o'},g_{o'},\phi_{o'})$ need not
converge to each other when $o' \rightarrow o$. However, $({\mathcal
  M}_{o'},g_{o'},\phi_{o'})$ will converge to some dynamical trajectory that is
Diff$_\infty{\mathcal M}_o$-equivalent to $({\mathcal M}_o,g_o,\phi_o)$ as $o'
\rightarrow o$ and we can include such diffeomorphisms into $\iota_o$'s to
make $\sigma$ differentiable.

The condition 1. and the discussion before it imply that $\sigma|_o$ is a
differentiable injection with differentiable inverse for any
$o\in\Gamma/o$. The condition 2.\ implies then that the maps $\pi$ and
$\sigma$ are transversal to each other. That is the following property. Let
$p$ be an arbitrary point of $\Gamma$. Then any vector $v\in T_p(\Gamma)$ that
satisfies both equations $d\sigma(v) = 0$ and $d\pi(v) = 0$ must be the zero
vector.

We shall call the pair $(\pi,\sigma)$ of maps a {\em Kucha\v{r}
  $\Gamma$-decomposition} (recall that $\pi$ is the projection defined in
Sec.\ \ref{sec:G}). Thus, for any $p\in \Gamma_\Sigma$, $\pi(p)$ is a point in
the physical phase space of the model, and $\sigma(p)$ is some embedding,
$\sigma(p) :\Sigma\mapsto\Sigma\times{\mathbf R}$, giving the position of
``the slice'' $\Sigma$ in ``the background manifold'' $\Sigma\times{\mathbf
  R}$. 

The image of $\Gamma_\Sigma$ under $\pi\times\sigma$ is a subset
$(\pi\times\sigma)(\Gamma_\Sigma)$ of the Cartesian product manifold $K_\Sigma
:= (\Gamma_\Sigma/o)\times{\mathrm Emb}(\Sigma,\Sigma\times{\mathbf R})$.
$\sigma(o)$ is some open subset of ${\mathrm Emb}(\Sigma,\Sigma\times{\mathbf
  R})$ for each $o$. These subsets can vary with varying
$o\in\Gamma_\Sigma/o$. Indeed, consider an embedding
$h:\Sigma\mapsto(\Sigma\times{\mathbf R})$ and study its pull-back
$h':\Sigma\mapsto{\mathcal M}_o$ by $\iota_o$ defined by $h' :=
\iota^{-1}_o\circ h$. Such a pull-back need not exist, because $\iota_o$ is
only injection so there may be points in $\Sigma\times{\mathbf R}$ that do not
lie in the range of $\iota_o$ and $h(\Sigma)$ can contain such points. Even if
the pull-back exists, it need not define a spacelike surface in $({\mathcal
  M}_o,g_o)$, and even if the surface is spacelike, it need not be a Cauchy
surface. For a different c-orbit $o'$, $o'\neq o$, the $\iota_{o'}$-pull-back
of the same embedding $h$ can have quite different properties as to its
existence and as to its being a Cauchy surface in $({\mathcal
  M}_{o'},g_{o'})$.  Thus, the projection of the subset
$(\pi\times\sigma)(\Gamma_\Sigma)$ of $K_\Sigma$ to ${\mathrm Emb}(\Sigma,
\Sigma\times{\mathbf R})$ may be empty; on the other hand, its projection onto
$\Gamma_\Sigma/o$ coincides with $\Gamma_\Sigma/o$.

From our definitions and assumptions, it follows that the map $\pi\times\sigma
: \Gamma_\Sigma \mapsto K_\Sigma$ is a differentiable injection with
differentiable inverse.  Indeed, it is an injection because all points that
have the same projection by the submersion $\pi$ form a c-orbit $o$, $o$ is
mapped by the bijection $\rho_o$ to ${\mathrm Emb}_c(\Sigma_o,{\mathcal
  M}_o)$, and this set by the injection $\tilde{\iota}_o$ induced by $\iota_o$
to $\sigma(o) \subset {\mathrm Emb}(\Sigma, \Sigma\times{\mathbf R})$ (of
course, $\Sigma$ is diffeomorphic to $\Sigma_o$). It is differentiable, because
both maps, $\pi$ and $\sigma$, are. It has a differentiable inverse because
$\pi$ and $\sigma$ are transversal.

Then there is a well-defined presymplectic form $\Omega_K$ on
$(\pi\times\sigma)(\Gamma_\Sigma)$, $\Omega_K := (\pi\times\sigma)_*\Omega$,
the push-forward by the map $\pi\times\sigma$ of the presymplectic form
$\Omega$ on $\Gamma_\Sigma$. It is easy to see what the structure of
$\Omega_K$ is: Its characteristic subspaces are tangential to ${\mathrm
  Emb}(\Sigma, \Sigma\times{\mathbf R})$ at any point
$\Bigl(\pi(p),\sigma(p)\Bigr)$, where $o$ is the c-orbit through $p$, $o =
\pi^{-1}\Bigl(\pi(p)\Bigr)$. Its pull-back to $\Gamma_\Sigma/o$ coincides with
the form $\tilde{\Omega}$ defined in Sec.\ \ref{sec:G}.

We can go further and consider the space $K_\Sigma$ to be a trivialization of
a fiber bundle $E_\Sigma$ with the base space $\Gamma/o$, the typical fiber
${\mathrm Emb}(\Sigma, \Sigma\times{\mathbf R})$ and the group
Diff$_\infty(\Sigma\times{\mathbf R})$. Each $\sigma$ can be decomposed into a
direct map $\kappa : \Gamma_\Sigma \mapsto E_\Sigma$, which is a
differentiable injection with differentiable inverse, a trivialization $\tau :
E_\Sigma \mapsto K_\Sigma$, and the projection $\eta : K_\Sigma \mapsto
{\mathrm Emb}(\Sigma, \Sigma\times{\mathbf R})$, so that $\sigma =
\eta\circ\tau\circ\kappa$.  $\kappa$ is independent of covariant gauge
fixings; each fixing, however, defines $\sigma$, and so a trivialization
$\tau$ of $E_\Sigma$.

The covariant gauge fixing determines also a unique set of fields and branes
with a domain in the background manifold $\Sigma\times{\mathbf R}$ for any
dynamical trajectory, that is for any c-orbit $o \in \Gamma/o$; let us denote
this set by $\Bigl(g(o),\phi(o)\Bigr)$, where $o\in\Gamma/o$. This can be seen
as follows.  For each orbit $o$, we have a definite representative $({\mathcal
  M}_o,g_o,\phi_o)$; the spacetime manifold ${\mathcal M}_o$ is mapped by the
diffeomorphism $\iota_o$ into the background manifold ${\mathcal M}$; hence,
we can define
\be
  g(o) := \iota_{o*}g_o,\quad \phi(o) := \iota_{o*}\phi_o,
\label{KRV}
\ee 
where $\iota_{o*}$ is the push-forward defined by the map $\iota_o$.
$g(o)$ is the coordinate-free version of the Kucha\v{r}, Romano and
Varadarajan metric\cite{KRV} mentioned in Sec.\ \ref{sec:gauge}.

Observe that the unique set $\Bigl(g(o),\phi(o)\Bigr)$ is exactly what one
expects a gauge fixing to deliver: a unique set of fields and branes on the
background manifold $\Sigma\times{\mathbf R}$ for each class $\{({\mathcal
  M}_o,g_o,\phi_o)\}$. In fact, the set $\Bigl(g(o),\phi(o)\Bigr)$ represents
what can be called a locally {\em complete solution} to the dynamical
equations: each dynamical trajectory of an open set is obtained by a suitable
choice of $o$.

In this section, we have deliberately left the map $\sigma$ completely general.
Of course, one can easily construct a map $\sigma$ if one knows a {\em
  complete} coordinate condition that works in a neighbourhood of a {\em
  whole} Cauchy surface and admits a sufficiently large set of initial data on
this surface to cover a {\em whole} open set $\tilde{U}$ in the physical phase
space. A ``complete'' condition defines a unique coordinate system in certain
domain in each maximal dynamical trajectory corresponding to a point of
$\tilde{U}$. For example, the harmonic coordinate condition is not complete in
this sense. Some conditions work even globally. An example for general
relativity coupled to special kind of continuous matter\cite{KSG,KM} has been
given. For pure gravity, suppose e.g.\ that all spatially compact maximal
solutions of Einstein's equations admit a unique and complete foliation by
surfaces $\Sigma_K$ of constant mean external curvature $K$, $K:=
q^{kl}K_{kl}$ (this is a form of the well-known CMC hypothesis). Then one can
stipulate that the surfaces $\Sigma_K$ are mapped onto $(\Sigma,K)$ in
$(\Sigma\times{\mathbf R})$ by each $\iota_o$ and hope that the map $\iota_o$
can be completed suitably inside of each $\Sigma_K$. Then the corresponding
covariant gauge fixing is also global. In general, it seems plausible that any
construction of a particular map $\sigma$ can be based on a set of
differentiable sections of the submersion $\pi$ on $\Gamma$.

The simplest construction of $\sigma$, however, would starts from a locally
complete solution $\Bigl({\mathcal M},g(o),\phi(o)\Bigr)$, if such is known.
Each $o\in\Gamma/o$ is determined by values of a suitable set of constants of
motion and $\pi$ is trivial. For each $o$ and each Cauchy embedding $h :
\Sigma \mapsto {\mathcal M}$, one then calculates the initial datum
$(\Sigma,\gamma,\psi,\dot{\gamma},\dot{\psi})$ that the fields and branes
$g(o)$ and $\phi(o)$ induce on $h(\Sigma)$. This defines $(\sigma_o)^{-1}$. We
shall use this construction in subsequent papers.

\section{Extensions of Kucha\v{r} decompositions from $\Gamma$ to $\Gamma'$}
\label{sec:ext}
A covariant gauge fixing described in the previous section defines a division
of variables into two groups: the set of dynamical variables that determine
points of the physical phase space $\Gamma/o$ and the set of kinematical
variables that describe an embedding of the Cauchy surface $\Sigma$ into a
background manifold $\mathcal M$; this division is done without use of
coordinates in any of these manifolds. It is, however, not yet a full
Kucha\v{r} decomposition as outlined in Sec.\ \ref{sec:exampl}, because it
works only inside the constraint surface $\Gamma$, whereas the original
Kucha\v{r} decomposition holds in a neighbourhood of $\Gamma$ in the phase
space $\Gamma'$. In the present section, we shall extend gauge fixings and
Kucha\v{r} decompositions from $\Gamma$ to $\Gamma'$. We shall work with a
fixed $\Sigma$-sector of the model, and we leave out the corresponding index
$\Sigma$ everywhere.

Let us first describe what exactly is the problem. Clearly the components of
the symplectic form $\Omega'$ on $\Gamma'$ with respect to Kucha\v{r}
coordinates are
\[
  \Omega' = \int_\Sigma d^3x\Bigl(d{\mathcal H}_\mu(x)\wedge dX^\mu(x) +
  dp_\alpha(x)\wedge dq_\alpha(x)\Bigr).
\]
This can be written as $\Omega' = \Omega_1 \oplus_\bot \Omega_2$, where
\[
  \Omega_1 := d\int_\Sigma d^3x\, p_\alpha(x)\wedge dq^\alpha(x)
\]
is a symplectic form on $\Gamma/o$ with coordinates $p_\alpha(x)$ and
$q^\alpha(x)$; in the form  
\be
  \Omega_2 := d\int_\Sigma d^3x\, {\mathcal H}_\mu(x)\wedge dX^\mu(x)
\label{o2}
\ee 
we clearly recognize the canonical symplectic structure of
$T^*\Bigl(\text{Emb}(\Sigma,\Sigma\times{\mathbf R})\Bigr)$, where $X^\mu(x)$
is a point of the manifold $\text{Emb}(\Sigma,\Sigma\times{\mathbf R})$ and
${\mathcal H}_\mu(x)$ is a cotangent vector at $X^\mu(x)$.

The set $(\pi\times\sigma)(\Gamma)$ is a submanifold of $K$, which is in turn
a submanifold of $K' := (\Gamma/o)\times
T^*\Bigl(\text{Emb}(\Sigma,\Sigma\times{\mathbf R})\Bigr)$ that is determined
by the equations ${\mathcal H}_\mu(x) = 0$ in $K'$. Hence, $\pi\times\sigma$
injects $\Gamma$ symplectically into $K'$. What we are looking for is,
therefore, a symplectic injection $\varphi$ that maps a neighbourhood $U'$ of
$\Gamma$ in $\Gamma'$ into $K'$ so that $\varphi|_\Gamma = \pi\times\sigma$.

We shall show the existence of such an extension $\varphi$ in three steps. The
proof will be given in a form that is immediately valid only for
finite-dimensional manifolds. After each step, however, we shall discuss the
points that do not admit a straightforward generalization to infinite
dimensional cases and show how the argument can be improved.

\subsection{Extension of $\pi\times\sigma$ to the tangent space of $\Gamma$ in
  $\Gamma'$} 
First, we extend $\pi\times\sigma$ just ``to the first order'' at
$\Gamma$, that is, we construct a map $\varphi_1 : T'(\Gamma) \mapsto T(K')$,
where $T'(\Gamma)$ is the vector bundle with the base space $\Gamma$ whose
fibers are the spaces $T_p(\Gamma')$ tangent to $\Gamma'$ at all $p\in
\Gamma$; it is a subbundle of $T(\Gamma')$ which could also be denoted by
$T_p(\Gamma')|_\Gamma$. The map $\varphi_1$ must have the following
properties: (i) $\varphi_1$ is a vector bundle morphism, (ii)
$\varphi_1|_{T(\Gamma)} = d(\pi\times\sigma)$, and (iii) $\varphi_1$ is
symplectic.  Because of (i), $\varphi_1$ can be decomposed\cite{lang} into a
set of maps containing a base-space map $\varphi_{1b} : \Gamma \mapsto K$, and
fiber maps $\varphi_{1fp} : T'(\Gamma) \mapsto T_{\varphi_{1b}(p)}(K')$ for
each $p \in \Gamma$; $\varphi_{1b}$ is a differentiable injection and
$\varphi_{1fp}$ is a linear isomorphism for each $p \in \Gamma$. Because of
(ii), $\varphi_{1b} = \pi\times\sigma$ and
\be
  \varphi_{1fp}|_{T_p(\Gamma)} = d(\pi\times\sigma)|_{T_p(\Gamma)}.
\label{18e.1}
\ee
Finally, because of (iii), $\varphi_{1fp}$ is a symplectic
isomorphism at each $p\in\Gamma$.

The map $\varphi_{1fp}$ is, therefore, already determined on the subspace
$T_p(\Gamma)$ of $T'_p(\Gamma)$, and we have to specify it only on a subspace,
say, $N_p(\Gamma)$ of $T'_p(\Gamma)$ such that $T'_p(\Gamma) = N_p(\Gamma)
\oplus T_p(\Gamma)$. 
 
The symplectic forms $\Omega'$ and the pull-back $\varphi^{
  *}_{1fp}\Omega_{K'}$ must, moreover, coincide on $T'_p(\Gamma)$ for all
$p\in\Gamma$; they do so already on $T_p(\Gamma)$. This suggests an idea for
the construction. The image $T_{\varphi_{1b}(p)}(K')$ splits in a way adapted
to the symplectic form $\Omega_{K'}$:
\[
  T_{\varphi_{1b}(p)}(K') = T_{\varphi_{1b}(p)}(K) 
  \oplus T_{\sigma(p)}
  \Biggl(T_{\sigma(p)}^*\Bigl(\text{Emb}(\Sigma, \Sigma\times{\mathbf
  R})\Bigr)\Biggr),   
\]
where the space $T_{\sigma(p)}
\Biggl(T_{\sigma(p)}^*\Bigl(\text{Emb}(\Sigma, \Sigma\times{\mathbf
  R})\Bigr)\Biggr)$ is $\Omega_{K'}$-isotropic (i.e., restriction of
$\Omega_{K'}$ to this space is a zero form) and $\Omega_{K'}$-orthogonal to
$T_{\pi(p)}(\Gamma/o)$. Hence, the pre-image $N_p(\Gamma)$ of this
space,
\be
  N_p(\Gamma) := \varphi^{-1}_{1fp}
  T_{\sigma(p)}\Biggl(T_{\sigma(p)}^*
  \Bigl(\text{Emb}(\Sigma,\Sigma\times{\mathbf R})\Bigr)\Biggr) 
\label{18f.1}
\ee
must be isotropic in $T'_p(\Gamma)$ with respect to $\Omega'$ and
$\Omega'$-orthogonal to $Q_p(\Gamma)$, which is the pre-image of
$T_{\pi(p)}(\Gamma/o)$, $Q_p(\Gamma) :=
\varphi^{-1}_{1fp}T_{\pi(p)}(\Gamma/o)$. However, $\Gamma/o \subset K$, so
$\varphi^{-1}_{1fp}$ on $\Gamma/o$ is $\Bigl(d(\pi\times\sigma)\Bigr)^{-1}$,
and we have finally 
\[
  Q_p(\Gamma) := \Bigl(d(\pi\times\sigma)\Bigr)^{-1}T_{\pi(p)}(\Gamma/o).
\]
$Q(\Gamma)$ must be a smooth vector bundle whose basis is the constraint
manifold; our construction starts from this bundle.

The crucial observation now is that any subspace $N_p(\Gamma)$ of
$T'_p(\Gamma)$ that satisfies (a) $T'_p(\Gamma) = N_p(\Gamma) \oplus
T_p(\Gamma)$, (b) $N_p(\Gamma)$ is $\Omega'$-orthogonal to $Q_p(\Gamma)$ and
(c) $N_p(\Gamma)$ is $\Omega'$-isotropic, defines a suitable symplectic map
$\varphi_{1fp}$ by the requirement (\ref{18f.1}): as
$\varphi_{1fp}$ is linear, and because of the condition (a), the
knowledge of $\varphi_{1fp}$ on $T_p(\Gamma)$ (which is well-known,
see Eq.\ (\ref{18e.1})) and on $N_p(\Gamma)$ determines
$\varphi_{1fp}$ uniquely.

As is already suggested by the notation, $N(\Gamma)$ is to be a smooth vector
bundle in order that $\varphi_1$ is a differentiable map. A construction of an
example of such an $N(\Gamma)$ would show the existence of $\varphi_1$.

The vector bundle $Q(\Gamma)$ is a subbundle of $T(\Gamma)$ which, in turn, is
a subbundle of $T'(\Gamma)$. As $\Gamma$ is a submanifold of $\Gamma'$, there
must be a vector bundle $N(\Gamma)$ such that
\be 
  T'(\Gamma) = N(\Gamma) \oplus T(\Gamma),
\label{18.1}
\ee 
where $N(\Gamma)$ is a (smooth) vector bundle. If $N(\Gamma)$ is isotropic and
orthogonal to $Q(\Gamma)$, then it is the desired bundle. If $N_p(\Gamma)$ is
not orthogonal to $Q_p(\Gamma)$, we can find a continuous linear map $\psi_Q :
N_p(\Gamma) \mapsto T_p(\Gamma')$ such that $\psi_Q\Bigl(N_p(\Gamma)\Bigr)$ is
orthogonal to $Q_p(\Gamma)$ and $T_p(\Gamma') = \psi_Q\Bigl(N_p(\Gamma)\Bigr)
\oplus T_p(\Gamma)$ as follows.

Recall that $L_p(\Gamma)$ is the characteristic subspace, Eq.\ (\ref{char}),
and that
\be
  T_p(\Gamma) = L_p(\Gamma) \oplus Q_p(\Gamma)
\label{18i.1}
\ee
for all $p\in\Gamma$, because $\pi$ and $\sigma$ are transversal to each
other.  It follows that $\Omega'$ must be non-degenerate on $Q_p(\Gamma)$.
Indeed, if $v\in Q_p(\Gamma)$ were $\Omega'$-orthogonal to all vectors of
$Q_p(\Gamma)$, then it would also be $\Omega'$-orthogonal to all of
$T_p(\Gamma)$ and so it would belong to $L_p(\Gamma)$. Then, however, $v=0$
because of Eq.\ (\ref{18i.1}).

If $\Omega'$ is non-degenerate on $Q_p(\Gamma$), then there is a unique vector
$a\in Q_p(\Gamma)$ for each linear function $\alpha$ on $Q_p(\Gamma)$ such
that $\alpha(v) = \Omega'(a,v)$ for all $v\in Q_p(\Gamma)$. Let $w\in
T'_p(\Gamma)$; then $\Omega'(w,\cdot)$ is a linear function on $Q_p(\Gamma)$
and it determines, therefore, a unique element $\omega_Q(w) \in Q_p(\Gamma)$
such that 
\be
  \Omega'(w,v) = \Omega'\Bigl(\omega_Q(w),v\Bigr)
\label{oqu}
\ee
for all $v\in Q_p(\Gamma)$; the map $\omega_Q : T'_p(\Gamma) \mapsto
Q_p(\Gamma)$ is linear. The desired map $\psi_Q$ is then defined by 
\be
  \psi_Q := {\mathrm
  id}_{N_p(\Gamma)} - \omega_Q|_{N_p(\Gamma)}.
\label{18j.1}
\ee

Orthogonality can be shown as follows. Let $n\in N_p(\Gamma)$, and let us
calculate $\Omega'\Bigl(\psi_Q(n),q\Bigr)$ for any $q\in Q_p(\Gamma)$:
\[
  \Omega'\biggl(n - \omega_Q(n),q\biggr) = \Omega'(n,q) -
  \Omega'(\omega_Q(n),q) =0
\]
because of Eq.\ (\ref{oqu}).  Moreover, any $v\in T'_p(\Gamma)$ can be written
as $v = n + t$, where $n\in N_p(\Gamma)$ and $t\in T_p(\Gamma)$; then we have
also
\be
  v = \psi_Q(n) + \biggl(t + \omega_Q(n)\biggr).
\label{20.1}
\ee
From the definition of $\omega_Q$, it follows that $\omega_Q(n)\in Q_p(\Gamma)$
for all $n\in N_p(\Gamma)$, so $\Bigl(t + \omega_Q(n)\Bigr) \in T_p(\Gamma)$
and $\psi_Q\Bigl(N_p(\Gamma)\Bigr) \cap T_p(\Gamma) = 0$. Then the
decomposition (\ref{20.1}) is unique and the property follows. Thus, we have
proved Eq.\ (\ref{18.1}) with $N_p(\Gamma)$ being everywhere orthogonal to
$Q_p(\Gamma)$.

If $N_p(\Gamma)$ is orthogonal but not isotropic, we can improve it further as
follows. Consider the space $Q_p^\bot(\Gamma)$ defined by $Q_p^\bot(\Gamma) :=
\text{orth}_{\Omega'}Q_p(\Gamma)$. Here, we denote the space of all vectors of
$T'_p(\Gamma)$ that are $\Omega'$-orthogonal to $Q_p(\Gamma)$ by
$\text{orth}_{\Omega'}Q_p(\Gamma)$. As $\Omega'|_{Q_p(\Gamma)}$ is not
degenerate, we must have
\be
  T'_p(\Gamma) = Q_p(\Gamma)\oplus Q^\bot_p(\Gamma).
\label{18k.1}
\ee
Indeed, $Q_p^\bot(\Gamma)$ can be constructed from any linear complement
$V_p(\Gamma)$ of $Q_p(\Gamma)$ in $T'_p(\Gamma)$ by $Q_p^\bot(\Gamma) =
\psi_Q(V_p(\Gamma))$, where $\psi_Q$ is defined by Eq.\ (\ref{18j.1}) and the
proof is analogous to that for $\psi_Q\Bigl(N_p(\Gamma)\Bigr)$.

The spaces $N_p(\Gamma)$ and $L_p(\Gamma)$ are subspaces of
$Q_p^\bot(\Gamma)$. They are disjoint, $N_p(\Gamma) \cap L_p(\Gamma) = \{0\}$,
because $N_p(\Gamma) \cap T_p(\Gamma) = \{0\}$ and $L_p(\Gamma)\subset
T_p(\Gamma)$. Moreover, if $v\in Q_p^\bot(\Gamma)$, then $v = n + t$, where $n
\in N_p(\Gamma)$ and $t\in T_p(\Gamma)$ and also $t = x + q$, where $x\in
L_p(\Gamma)$ and $q\in Q_p(\Gamma)$ because of Eq.\ (\ref{18i.1}). Thus, we
obtain that $v = n + x + q$. Now, $v\in Q_p^\bot(\Gamma)$, $n\in
Q_p^\bot(\Gamma)$, and $x\in Q_p^\bot(\Gamma)$, hence also $q\in
Q_p^\bot(\Gamma)$, and as $Q_p^\bot(\Gamma)\cap Q_p(\Gamma) = \{0\}$, we must
have $q = 0$, so $v = n + x$. We have thus shown that
\be 
  Q_p^\bot(\Gamma) = N_p(\Gamma) \oplus L_p(\Gamma).
\label{23.1}
\ee
From the definition (\ref{char}) of $L_p(\Gamma)$ it follows that each vector
of $Q^\bot_p(\Gamma)$ that is $\Omega'$-orthogonal to all of $L_p(\Gamma)$
must lie in $L_p(\Gamma)$. For such a vector is $\Omega'$-orthogonal to both
$L_p(\Gamma)$ and $Q_p(\Gamma)$, and so to all of $T_p(\Gamma)$ because of
Eq.\ (\ref{18i.1}).

Now suppose that $\alpha : N_p(\Gamma) \mapsto {\mathbf R}$ is any linear
function on $N_p(\Gamma)$. We can extend such a function to $\bar{\alpha} :
Q^\bot_p(\Gamma)\mapsto {\mathbf R}$ by requiring
\be
  \bar{\alpha}|_{L_p(\Gamma)} = 0
\label{18m.1}
\ee
As $\Omega'$ is non-degenerate on $Q^\bot_p(\Gamma)$, there is a unique vector
$b\in Q^\bot_p(\Gamma)$ such that $\bar{\alpha}(u) = \Omega'(b,u)$ for all
$u\in Q^\bot_p(\Gamma)$. However, such a vector $b$ must then lie in
$L_p(\Gamma)$ because of Eq.\ (\ref{18m.1}). Hence: for any linear function
$\alpha$ on $N_p(\Gamma)$, there is $l\in L_p(\Gamma)$ such that $\alpha(u) =
\Omega'(l,u)$ for all $u\in Q^\bot_p(\Gamma)$. 

Let $n\in N_p(\Gamma)$; then $\Omega'(v,n)$ can be considered as a linear
function on $N_p(\Gamma)$ for any $v\in T'_p(\Gamma)$. There is, therefore, a
unique element $\omega_L(v)$ of $L_p(\Gamma)$ such that
\be
  \Omega'(v,n) = \Omega'(\omega_L(v),n).
\label{18n.1}
\ee
From its construction, it follows that $\omega_L : T'_p(\Gamma) \mapsto
L_p(\Gamma)$ is a linear map.

Consider the linear map $\psi_L : N_p(\Gamma) \mapsto T'_p(\Gamma)$ defined by
\[
  \psi_L(n) := n - (1/2)\omega_L(n)
\]
for all $n\in N_p(\Gamma)$. $\psi_L$ is an injection, because the equation $n
= (1/2)\omega_L(n)$ can have only zero solutions. Indeed, the left-hand side
is an element of $N_p(\Gamma)$ and the right-hand side is an element of
$L_p(\Gamma)$. Moreover, the image, $\psi_L\Bigl(N_p(\Gamma)\Bigr)$ is
isotropic; this can be seen as follows. Using the definition of  $\psi_L$ we
obtain 
\begin{eqnarray*}
  \lefteqn{\Omega'(\psi_L(n_1),\psi_L(n_2)) = } \\
  && \Omega'(n_1,n_2) - (1/2)\Omega'\Bigl(\omega_L(n_1),n_2\Bigr) +
  (1/2)\Omega'\Bigl(\omega_L(n_2),n_1\Bigr) + 
  (1/4)\Omega'\Bigl(\omega_L(n_1),\omega_L(n_2)\Bigr).
\end{eqnarray*}
The last term is zero, because $\omega_L(n_{1,2})\in L_p(\Gamma)$ and we
finally have from Eq.\ (\ref{18n.1})
\[
 \Omega'(\psi_L(n_1),\psi_L(n_2)) = \Omega'(n_1,n_2) -
  (1/2)\Omega'(n_1,n_2) + (1/2)\Omega'(n_2,n_1) = 0.
\]  

The last property we need is that any vector $q^\bot\in Q^\bot_p(\Gamma)$ can
be written as a sum, $q^\bot = z + y$, where
$z\in\psi_L\Bigl(N_p(\Gamma)\Bigr)$ and $y\in L_p(\Gamma)$. However, it holds
that $q^\bot = n + x$, where $n\in N_p(\Gamma)$ and $x\in L_p(\Gamma)$. Then
$q^\bot = \Bigl(n - (1/2)\omega_L(n)\Bigr) + \Bigl(x + (1/2)\omega_L(n)\Bigr)$
so $z = \psi_L(n)$ and $y = x + \omega_L(n)$ is the desired decomposition.

The restriction of $\varphi_{1fp}$ to $N_p(\Gamma)$ is uniquely determined by
the condition that $\varphi_{1fp} : T'_p(\Gamma) \mapsto
T_{\varphi_{1b}(p)}(K')$ is symplectic as follows. Let $n\in N_p(\Gamma)$,
then there is a unique linear function $\Omega'(n,\cdot)$ on $L_p(\Gamma)$
defined by $n$. $L_p(\Gamma)$ is mapped by $\varphi_{1fp}|_{L_p(\Gamma)} =
d\sigma$ onto $T_{\sigma(p)}\Bigl(\text{Emb}(\Sigma,\Sigma\times{\mathbf
  R})\Bigr)$, so $\Omega'(n,d\sigma^{-1}\cdot)$ is a linear function on
$T_{\sigma(p)} \Bigl(\text{Emb}(\Sigma,\Sigma\times{\mathbf R})\Bigr)$. Every
linear function $\alpha$ on
$T_{\sigma(p)}\Bigl(\text{Emb}(\Sigma,\Sigma\times{\mathbf R}) \Bigr)$
determines, in turn, a unique element $v$ of
$T_{\sigma(p)}\Biggl(T^*_{\sigma(p)}
\Bigl(\text{Emb}(\Sigma,\Sigma\times{\mathbf R})\Bigr)\Biggr) =
\varphi_{1fp}N_p(\Gamma)$ such that $\Omega'(v,\cdot)) = \alpha$. We must set
\[
  \varphi_{1fp}(n) := v,
\]
or else $\varphi_{1fp}$ is not symplectic. Hence, the choice of the subspace
$N_p(\Gamma)$ determines $\varphi_{1fp}$.

The above construction of $N_p(\Gamma)$ that satisfies the requirements (a),
(b) and (c) is based on the smooth vector bundles $Q(\Gamma)$ and $L(\Gamma)$
and on the differentiable symplectic form $\Omega'$; the result is, therefore,
a smooth vector bundle $N(\Gamma)$ .

In the case that the space $T'_p(\Gamma)$ is an infinite-dimensional Banach
space, two aspects of our construction become problematical.
\begin{enumerate}
\item If a Banach space $B_1$ is, as a linear space, a direct sum of two other
  linear (Banach) spaces, $B_1 = B_2 \oplus B_3$, then the corresponding map
  between $B_1$ and $B_2\times B_3$ need not be a {\em topological}
  isomorphism. If it is, one says that $B_2$ {\em splits}\cite{lang} $B_1$.
  Some splittings follow from the assumptions in subsections \ref{sec:E} and
  \ref{sec:G}. For example, Eq.\ (\ref{18.1}) follows from $\Gamma$ being a
  submanifold of $\Gamma'$, and Eq.\ (\ref{18i.1}) follows from $\pi$ being a
  submersion\cite{lang}. We however also need Eqs.\ (\ref{18k.1}) and
  (\ref{23.1}).
\item The norm defining $T'_p(\Gamma)$ can restrict the elements of
  $T'_p(\Gamma)$ so much that the space of all continuous linear functionals
  on $T'_p(\Gamma)$---the dual space $T^{\prime *}_p(\Gamma)$---contains also
  functionals that are not of the form $\Omega'(u,\cdot)$ for $u\in
  T'_p(\Gamma)$, even if $\Omega'(u,\cdot)$ is a non-zero functional for all
  non-zero $u$'s. Such a $\Omega'$ is called a {\em weakly non-degenerate} or
  {\em weak} symplectic form\cite{C-M}. If $\Omega'$ is weak, the definition
  of the maps $\omega_Q$ and $\omega_L$ given above does not work.
\end{enumerate}

The way one can cope with these two problems depends on the topology of
$T'_p(\Gamma)$. This, however, must be judiciously adapted to the nature of
each particular model and there does not seem to be any general method. Still,
the following scheme has worked for all examples we have considered as yet.
First, one defines certain dense subspaces, $T'_{p0}(\Gamma)$ and $T^{\prime
  *}_{p0}(\Gamma)$, of the Banach spaces $T'_{p}(\Gamma)$ and $T^{\prime
  *}_{p}(\Gamma)$; one can take, for example, the spaces of functions all of
whose derivatives are smooth and which have compact support. One has to prove
that $Q_p(\Gamma)\cap T'_{p0}(\Gamma)$ and $L_p(\Gamma)\cap T'_{p0}(\Gamma)$
are also dense in $Q_p(\Gamma)$ and $L_p(\Gamma)$, and that all functionals
from $T^{\prime *}_{p0}(\Gamma)$ have the form $\Omega'(u,\cdot)$ where $u\in
T'_{p0}(\Gamma)$. Then the construction seems to work for the corresponding
dense subspaces such as $Q_{p0}^\bot(\Gamma)$ or $N_{p0}(\Gamma)$---the
topology is nowhere needed. Second, one has to show that the projectors onto
the subspaces are continuous with respect to the topology. Then the Banach
spaces are easily shown to split and everything works.

\subsection{The pull-back of $\Omega_{K'}$ to $\Gamma'$}
The second step consists of two consecutive pull-backs that 
bring the form $\Omega_{K'}$ to $\Gamma'$. 

The restriction to the vector bundle $N(\Gamma)$ of the map $\varphi_1$
constructed in the previous subsection maps $N(\Gamma)$ to the vector bundle
with the base space $\varphi_{1b}\Gamma \subset K$, and with the fibers
$T_{\sigma(p)}\Biggl(T^*_{\sigma(p)}\Bigl(\text{Emb}(\Sigma,
\Sigma\times{\mathbf R})\Bigr)\Biggr)$,
\[
  \varphi_{1fp}N_p(\Gamma) =
  T_{\sigma(p)}\Biggl(T^*_{\sigma(p)}\Bigl(\text{Emb}
  (\Sigma,\Sigma\times{\mathbf R})\Bigr)\Biggr). 
\]
The cotangent space $T^*_{\sigma(p)}\Bigl(\text{Emb}
(\Sigma,\Sigma\times{\mathbf R})\Bigr)$ is a linear space, so it can be
identified with its tangent space at its zero vector. Hence, with this
identification, $\varphi_1|_{N(\Gamma)}$ can be considered as a bundle
morphism mapping $N(\Gamma)$ onto the bundle with the basis
$\varphi_{1b}\Gamma \subset K$, and the fibers
$T^*_{\sigma(p)}\Bigl(\text{Emb}(\Sigma,\Sigma\times{\mathbf R})\Bigr)$.
However, this vector bundle is nothing but a subbundle of $K' = (\Gamma/o)
\times T^*\Bigl(\text{Emb}(\Sigma,\Sigma\times{\mathbf R})\Bigr)$.  In this
way, using the map $\varphi_1$, we have constructed a bundle morphism between
$N(\Gamma)$ and $K'$.

Let us denote this morphism by $\varphi'_1 : N(\Gamma) \mapsto K'$ and the
pull-back of $\Omega_{K'}$ by $\varphi'_1$ to $N(\Gamma)$ by $\Omega_1$,
$\Omega_1 := \varphi^{\prime *}_1\Omega_{K'}$. $\Omega_1$ is a symplectic form
on the manifold $N(\Gamma)$; that is, $\Omega_1$ is a bilinear form in the
tangent space $T_P\Bigl(N(\Gamma)\Bigr)$ at each point $P\in N(\Gamma)$. Let
$P=p\in \Gamma$, then $T_p\Bigl(N(\Gamma)\Bigr)$ can be decomposed as follows:
\[
  T_p\Bigl(N(\Gamma)\Bigr) = T_p(\Gamma) \oplus T_p\Bigl(N_p(\Gamma)\Bigr).
\]
Again, $T_p\Bigl(N_p(\Gamma)\Bigr)$ can be identified with $N_p(\Gamma)$, so
by Eq.\ (\ref{18.1}) $T_p\Bigl(N(\Gamma)\Bigr) = T'_p(\Gamma)$. From the
construction of the map $\varphi'_1$, it follows that $\Omega'$ coincides with
$\Omega_1$ at $T'_p(\Gamma)$ for each $p\in \Gamma$, so we can write
\be 
  \Omega_1|_\Gamma = \Omega'|_\Gamma.
\label{25;1}
\ee
Observe that $\Omega_1$ is a kind of ``constant'' extension of
$\Omega'|_\Gamma$ to the whole bundle $N(\Gamma)$.

The construction of our second pull-back uses the theorem about the existence
of tubular neighbourhoods\cite{zielon,lang}. A tubular neighbourhood is a
generalization of the well-known notion of normal coordinate ball. In the case
we consider, the theorem states that there is a diffeomorphism $\varphi_2 :
U_1 \mapsto U_2$, where $U_1$ is a neighbourhood of the zero section $\Gamma$
in $N(\Gamma)$ and $U_2$ a neighbourhood of $\Gamma$ in $\Gamma'$ such that
$d\varphi_2|_\Gamma = {\mathrm id}$.

It follows that the pull-back $\Omega_2 := \varphi^{-1*}_2\Omega_1$ is a
symplectic form on $U_2$. As $d\varphi_2|_\Gamma$ is an identity, we have
\be
  \Omega_2|_\Gamma = \Omega_1|_\Gamma.
\label{23-1}
\ee

Most constructions of this subsection work for infinite dimensions, if we just
replace the word ``non-degenerate'' by ``weakly non-degenerate''
everywhere---the difference is not important here. All necessary splittings
can easily be shown. The construction of the tubular neighbourhood is mostly
straightforward, too. However, if a complicated set has been deleted from
$\Gamma'_1$, then one has to use a smooth partition of unity\cite{lang} and
it need not be trivial to show its existence. The proof will depend on the
properties of the particular model.

\subsection{Application of the Darboux-Weinstein theorem}
This step is an application of the Darboux-Weinstein
theorem\cite{wein,zielon}. This is a generalization of the well-known Darboux
theorem saying roughly that if two symplectic forms $\Omega_a$ and $\Omega_b$
on a manifold $\mathcal M$ coincide at a submanifold ${\mathcal N} \subset
{\mathcal M}$, then there is a diffeomorphism $\lambda : {\mathcal M} \mapsto
{\mathcal M}$ that, together with its derivative $d\lambda$, is trivial at
${\mathcal N}$, and that $\lambda^*\Omega_2 = \Omega_1$ in a neighbourhood of
${\mathcal N}$ in ${\mathcal M}$.

Consider the two forms $\Omega'$ and $\Omega_2$ in $U_2$.  Eqs.\ (\ref{23-1})
and (\ref{25;1}) imply that $\Omega'|_\Gamma = \Omega_2|_\Gamma$, so the
conditions of the Darboux-Weinstein theorem are satisfied. There is, therefore,
a diffeomorphism $\varphi_3 : U' \mapsto U_3$ of a neighbourhood $U'$ of
$\Gamma$ in $\Gamma'$ with $U_3 \subset U_2$
such that $\varphi^*_3\Omega_2 = \Omega'$.

Let us finally define the map $\varphi$ by 
\[
  \varphi := \varphi'_1\circ\varphi^{-1}_2\circ\varphi_3;
\]
it maps the neighbourhood $U'$ of $\Gamma$ in $\Gamma'$ onto the neighbourhood
$\varphi U'$ of $K$ in $K'$.  From the constructions above, it follows
immediately that $\varphi^*\Omega_{K'} = \Omega'$. The maps $\varphi_2$ and
$\varphi_3$ are identities if restricted to $\Gamma$ and the restriction of
$\varphi'_1$ to $\Gamma$ is $\pi\times\sigma$. Thus, $\varphi|_\Gamma =
\pi\times\sigma$ and the existence of the extension is shown.

The constructions of this subsection need considerable modification in the
case of infinite dimensions. Indeed, the Darboux-Weinstein theorem does not
hold for general weak symplectic forms\cite{Mcontra}. Marsden has, however,
proved an analogous theorem\cite{M} for weak symplectic forms if certain
additional conditions are imposed, and the conditions are chosen in such a
way that most models met in practice satisfy them. Thus, the modification
consists of a proof that the particular model under study satisfies the
assumptions of Marsden theorem.

The extension constructed in this section is not unique. Already the Step 1
was not unique, because the subspace $N_p(\Gamma)$ is determined only up to a
symmetric linear map\cite{zielon} on $L_p(\Gamma)$. The tubular neighbourhood
of the Step 2 is also quite arbitrary. Finally, the Darboux-Weinstein theorem
guarantees just the existence of $\varphi_3$, but it says nothing about its
uniqueness.

\section{Conclusions}  
We have defined a covariant gauge fixing as pointwise identification of
different solution spacetimes with each other so that a fixed background
manifold has resulted and the dynamics has been reduced to a field dynamics on
it. The fixing has first been defined on the constraint manifold of the
system; there are very many ways to choose it at least locally; different
gauge fixings are related by elements of the huge Bergmann-Komar group.

We have found a connection between covariant gauge fixings and Kucha\v{r}
$\Gamma$-decompositions of the constraint manifold: for any fixing, there is
exactly one decomposition. The decomposition itself amounts to a particular
choice of (local) foliation of the constraint manifold that is transversal to
the c-orbits.

Finally, we have shown that any Kucha\v{r} decomposition of the constraint
surface can be extended to a whole neighbourhood of the constraint surface.
This extension is not unique. In this way, the full Kucha\v{r} decomposition
is doubly non-unique: there are as many $\Gamma$-decompositions as covariant
gauge fixings, and each $\Gamma$-decomposition has many extensions. However,
the form of kinematic term of the Kucha\v{r} action (\ref{kuch-a}) is always
the same, the only interesting and nontrivial part being the algebra of the
observables, if we allow for more general\cite{paris} algebra than the
Heisenberg algebra of $q^\alpha_0$ and $p_{\alpha 0}$ in Eq.\ (\ref{kuch-a}).
The usefulness of the decomposition is based on the enormous simplification it
brings about in the description of generally covariant systems.

We would like to make two additional remarks. First, the structure of the weak
symplectic manifold $(K',\Omega')$ is typical for the so-called {\em already
  parametrized theories} such as a parametrized scalar field in flat
spacetime (see, e.g.\ Kucha\v{r}\cite{kuchbanf}). Our construction shows that
the generally covariant models are, in general, {\em not} already parametrized
theories for two quite different reasons. 1) We can prove that only always a
part of the symplectic manifold of the system has the structure
$(K',\Omega')$, namely just a sector corresponding to a fixed Cauchy surface.
Moreover, we have to exclude points in the constraint surface that correspond
to dynamical trajectories admitting any symmetry. In fact, Torre\cite{torre}
has shown that general relativity cannot be considered as already parametrized
theory the obstruction coming from points at the constraint surface $\Gamma$
that represent Cauchy data for spacetimes with Killing vectors; these points
are also excised in our paper. 2) For each subsystem that {\em is} equivalent
to an already parametrized system, such an equivalence is not {\em unique}.
There is one Kucha\v{r} $\Gamma$-decomposition $\pi\times\sigma$ for each
covariant gauge fixing, and there are many different, gauge dependent,
background manifolds. This is in stark contrast to the structure of an already
parametrized system such as in Kucha\v{r}\cite{kuchbanf}, where there is a
unique background manifold. The points of this manifold are defined by the
fixed background metric---the Minkowski metric on it. The constraint manifold
of a generally covariant model is just a bundle with {\em many} different
trivializations, unlike that of an already parametrized model, which is a
{\em unique} cartesian product.

Second, we observe that our construction is closely related to the problem of
the so-called abelianization of constraints\cite{H-T}. Indeed, the new
constraints given by the theorem can be taken as components ${\mathcal
  H}_\mu(x)$ of the cotangent vectors in $T^*\Bigl({\mathrm
  Emb}(\Sigma,\Sigma\times{\mathbf R})\Bigr)$ with respect to some coordinates
on $\Sigma$ and on ${\mathcal M} = \Sigma\times{\mathbf R}$. All these
``functions'' have vanishing Poisson brackets with each other. Of course, a
complete system of abelian constraint functions still need not exist, because
there need not be global coordinates on $\Sigma$ and $\mathcal M$, and the
points with symmetries are also excluded.

\section{Acknowledgments}
Important discussions with Ji\v{r}\'{\i} Bi\v{c}\'{a}k, Arthur Fischer, Helmut
Friedrich, Mark Gotay, Karel V. Kucha\v{r}, Vincent Moncrief and John T.
Whelan are acknowledged. This work was supported in part by the Swiss
Nationalfonds, and by the Tomalla Foundation, Zurich.


\begin{references}
\bibitem{berg}Bergmann, P. G., and Komar, A. B., Int.\ J. Theor.\ Phys.\ {\bf
     5}, 15 (1972).  \bibitem{hole}Einstein, A. and Grossmannn, M.,
   Zeitschrift f\"{u}r Mathematik und Physik {\bf 63}, 215 (1914).
\bibitem{F-H}Fredenhagen, K. and Haag, R., Commun.\ Math.\ Phys.\ {\bf 108}, 91
   (1987).  
\bibitem{stach}Stachel,J., ``Einstein's Search for General
   Covariance, 1912-1915'' in {\em Einstein and the History of General
   Relativity}, ed.\ by Howard, D. and Stachel, J.,  Birkh\"{a}user, Boston,
   1989.  
\bibitem{salamanca}Isham, C. J. in {\em Integrable Systems, Quantum
   Groups, and Quantum Field Theories}, Kluver Academic Publishers, London,
   1993.  
\bibitem{CBYA}Chocquet-Bruhat, Y., York, jr.\, J. W., and Anderson,
 A., {\em Curvature-Based Hyperbolic System for General Relativity}, Preprint,
 gr-qc/9802027.
\bibitem{paris}H\'{a}j\'{\i}\v{c}ek, P., {\em Choice of Gauge in Quantum
     Gravity}, talk at 19th Texas Symposium of Relativistic Astrophysics and
     Cosmology, Paris, Dec.\ 1998. Preprint, gr-qc/9903089.  
\bibitem{thirdway} Kucha\v{r}, K. V., J. Math.\ Phys.\ {\bf 13}, 768 (1972). 
\bibitem{cylindr}Kucha\v{r}, K. V., Phys.\ Rev.\ {\bf D4}, 955 (1971).
\bibitem{wein}Weinstein, A., Advances in Math.\ {\bf 6}, 329 (1971).
\bibitem{schwarz}Kucha\v{r}, K. V., Phys.\ Rev.\ {\bf D50}, 3961 (1994).
\bibitem{v-bundl}Abraham, R., Marsden, J. E., and Ratiu, T., {\em Manifolds,
     Tensor Analysis, and Applications}, Springer, New York, 1988.
\bibitem{zielon}Libermann, C. G., and Marle, C.-M., {\em Symplectic Geometry
    and Analytical Mechanics}, Reidel, Dordrecht, 1987.
\bibitem{lang}Lang, S., {\em Differential and Riemannian Manifolds}, Springer,
     Berlin, 1995.
\bibitem{M}Marsden, J. E., {\em Lectures on Geometrical Methods in
     Mathematical Physics}, SIAM, Philadelphia, 1989.
\bibitem{DW}DeWitt, B. C. in {\em Relativity, Groups, and Topology}, Vol.\ I,
     ed.\ by DeWitt, C. and DeWitt, B. C., Gordon and Breach, New York, 1964.
  \bibitem{canada}Kucha\v{r}, K. V., in {\em Proceedings of the 4th Canadian
     Conference on General Relativity and Relativistic astrophysics}, World
     Scientific, Singapore, 1992. 
\bibitem{goa}Kucha\v{r}, K. V., in {\em Highlights in Gravitation and
     Cosmology}, ed.\ by B.~R.~Iyer et al., Cambridge University Press,
     Cambridge, 1988.
\bibitem{ADM}Arnowitt, R. S., Deser, S., and Misner, C. W., in {\em
     Gravitation: An Introduction to Current Research}, Wiley, New York, 1962.
\bibitem{kijow}Kijowski, J., {\em Metody geometryczne w klasycznej mechnice i
     teorii pola} (Lectures on differential geometrical methods in classical
     mechanics and field theory), Warsaw University Press, Warsaw, 1977. 
\bibitem{KRV} Kucha\v{r}, K. V., Romano, J. and Varadarajan, Phys.\ Rev.\ {\bf
     D55}, 795 (1997). 
\bibitem{carlip}H\'{a}j\'{\i}\v{c}ek, P., J. Math.\ Phys.\ {\bf 39}, 4824
     (1998).   
\bibitem{PH1}H\'{a}j\'{\i}\v{c}ek, P., Phys.\ Rev.\ {\bf D57}, 936 (1998).
\bibitem{foot}In the present paper, we focus on the group of diffeomorphisms
  of four-dimensional (spacetime) manifolds. One often finds formalisms based
  on a different group, Diff$\Sigma$, for instance Fischer\cite{fisch}; it is
  also this group of three-dimensional diffeomorphisms that play a crucial
  role in the Ashtekar\cite{asht} approach to quantum gravity. Our approach
     includes the symmetry with respect to Diff$\Sigma$. 
\bibitem{C-G}Choquet-Bruhat, Y., and Geroch, R. P., Commun.\ Math.\ Phys.\
     {\bf14}, 329 (1969).  
\bibitem{ger}Geroch, R., P., J. Math.\ Phys.\ {\bf 11}, 437 (1970).
\bibitem{F-M}Fischer, A. E., and Marsden, J. E., in {\em General
     Relativity. An Einstein Centenary Survey}, ed.\ by S. W. Hawking and
     W. Israel, Cambridge University Press, Cambridge, 1979.
\bibitem{K-T}Kijowski, J. and Tulczyjew, W. M., {\em A Symplectic Framework
     for Field Theories}, Lecture Notes in Physics Vol.\ 107, Springer, Berlin,
     1979.
\bibitem{F-Mo}Fischer, A. E., and Moncrief, V., Gen.\ Rel.\ Grav.\ {\bf 28},
     207 (1996).
\bibitem{I-K}Isham, C. J., and Kucha\v{r}, K. V., Ann.\ Phys.\ {\bf 164}, 288
  (1985).
\bibitem{mon-dif}Moncrief, V., J. Math.\ Phys.\ {\bf 16}, 493 (1975). 
\bibitem{KSG}Kijowski, J., Sm\'{o}lski, A. and G\'{o}rnicka, A., Phys.\ Rev.\
     {\bf D41}, 1875 (1990).
\bibitem{KM}Kijowski, J. and Magli, G., Class.\ Quantum Grav.\ {\bf 15}, 3891
     (1998).  
\bibitem{C-M}Chernoff, P. R., and Marsden, J. E., {\em Properties
    of Infinite Dimensional Hamiltonian Systems}, Springer, Berlin, 1974.
\bibitem{Mcontra} Marsden, J. E., Proc.\ Amer.\ Math.\ Soc.\ {\bf 32}, 590
     (1972). 
\bibitem{kuchbanf}Kucha\v{r}, K. V. in {\em Relativity, Astrophysics, and
    Cosmology}, ed.\ by W. Israel, Reidel, Dordrecht, 1973.
\bibitem{torre}Torre, C. G., Phys.\ Rev.\ {\bf D46}, R3231 (1992).
\bibitem{H-T}Henneaux, M., and Teitelboim, C., {\em Quantization of Gauge
    Systems}, Princeton University Press, Princeton, 1992. P. 113.
\bibitem{fisch}Fischer, A. E., in {\em Relativity}, ed.\ Carmeli, M., Fickler,
     S. I., and Witten, L., Plenum, New York, 1970.
\bibitem{asht} Ashtekar, A., {\em Lectures on Non-Perturbative Canonical
     Gravity}, World Scientific, Singapore, 1991.
 \end{references}
\end{document}